\documentclass[doublecol]{epl2}

\usepackage{amsfonts}
\usepackage{amssymb}
\usepackage{mathrsfs}
\usepackage{latexsym}
\usepackage{amsmath}

\usepackage[utf8]{inputenc}
\usepackage{graphicx}

\usepackage[pdfstartview=FitH]{hyperref}

\title{One-dimensional tunneling of the two-body bound state}
\shorttitle{One-dimensional tunneling of a bound state}

\author{N.~Shypka\inst{1}\footnote{Present address: Department of Information Control Systems, Kharkiv National University of Radio Electronics (NURE), 14 Nauky Ave., Kharkiv, Ukraine} \and O.~Hryhorchak\inst{1} \and V.~Pastukhov\inst{1}}
\shortauthor{N. Shypka \etal}

\institute{
  \inst{1} Professor Ivan Vakarchuk Department for Theoretical Physics,
  Ivan Franko National University of Lviv, 12 Drahomanov Street, Lviv, Ukraine\\
  E-mail: natalya.shypka@gmail.com (N. Shypka);
  hrorest@gmail.com (O. Hryhorchak);
  volodyapastukhov@gmail.com (V. Pastukhov)
}


\abstract{
We consider bound and scattering states of the one-dimensional dimer formed by two coupled non-identical atoms when one of them also interacts with the zero-range potential located at the origin. By calculating the dimer localized and scattering wave functions, we identify properties of the system without the two-body bound-state breakdown. In general, we predict an enhancement of the dimer reflection compared to a single atom, except for a narrow region on the attractive side of the external potential.}

\begin{document}

\maketitle

\section{Introduction}
Success in solving the one-body problems 100 years ago convinced the scientific community to give a chance to the then-new and very unusual theory known now as quantum mechanics. These few simplified, but exactly solvable models of reality, predetermined the future development of physics, forming our understanding of the atomic and molecular properties and the quantum tunneling effect. 

Along these lines, the simplest model of the composite particle (bound state of an arbitrary number of atoms) scattering was proposed in \cite{Zakhariev_1964}, with the interaction between atoms modeled by an infinite square well. Later, a slightly modified model was reinvented and numerically studied in detail by Saito and Kayanuma \cite{Saito_1994}. These were the first works pointing out the resonant tunneling of the bound state. The topic has developed so much that review articles \cite{Bertulani_2015} and whole book chapters \cite{Razavy} are dedicated to it. Within simplified approaches such as the adiabatic
approximation \cite{Bertulani_2007} or the rigid-molecule model \cite{Krassovitskiy_2014}, the composite structure of the
particle leads to an effective potential barrier for the center-of-mass motion that is substantially suppressed compared to its bare single-particle counterpart, thereby enhancing transmittance \cite{Bacca_2006}. A similar conclusion was drawn \cite{Takigawa_1984} on the semiclassical level for the system with the internal oscillator degrees of freedom. A more natural choice for interaction, particularly for molecules, is the harmonic potential, which possesses an infinite number of bound states. Tunneling properties of such systems without \cite{Penkov_2000,Sato_2002,Goodvin_2005,Ahsan_2010} and with transitions to excited states \cite{Lee_2006,Hnybida_2008} were extensively studied even for an arbitrary number \cite{Gusev_2014,Shegelski_2019} of harmonically-coupled particles. The effects of continuous spectrum in the relative motion of two atoms were investigated in \cite{Shegelski_2008}. For such systems, a composite particle incident upon an external potential can dissociate into two atoms if its energy exceeds the threshold. The reverse situation \cite{Shegelski_2008_2}, when two atoms incorporate into a molecule, is also possible.

In the context of ultra-cold atomic gases where the effective range of the two-body potential is small compared to the characteristic scattering lengths and average distances between particles, the singular $\delta$-like (pseudo)potentials are typically adopted. Although in higher dimensions, contact interaction requires the renormalization (see, for instance, \cite{Hryhorchak_2023} for more details); it possesses a few essential features of real potentials, namely a single bound state formation and modification of the $s$-wave scattering states. {The hard-wall reflection of such a bound state was extensively studied \cite{Lee_2011, Pine_2013} in arbitrary dimension.} In this article, we consider the one-dimensional two-particle states (both bound and scattering) with negative total energies, modeling the interparticle attractive potential and the scattering barrier by $\delta$-functions. It is natural for the low-dimensional ultracold-atom setup (see \cite{Sowinski_2019,Mistakidis_2023} for review) that allows one to observe two-body \cite{Zurn_2012} and even few-body \cite{Zurn_2013} tunneling experimentally. This can be viewed as the escape dynamics of coupled atoms in the combined optical and magnetic potential. Such a few-body evolution, for which the Wentzel-Kramers-Brillouin method is not a reliable approximation \cite{Gharashi_2015}, was studied by means of a full numerically exact solution of the Schr\"odinger equation \cite{Lode_2012}, using a specific basis \cite{Lundmark_2015} to optimize the numerical procedure. Theoretically, the problem was addressed within the quasiparticle wave function approach \cite{Rontani_2012} and by a powerful technique for treating one-dimensional bosons and fermions at strong coupling elaborated in \cite{Volosniev_2014}. Further developments were mainly concerned with the trap anharmonicity \cite{Ishmukhamedov_2017} and tunneling from the excited states \cite{Ishmukhamedov_2019}. Absence of the Pauli exclusion for bosons greatly facilitates the experimental observation of the many-body \cite{Zhao_2017} tunneling dynamics. Escape properties of the one-dimensional few-boson systems interacting through the $\delta$-potential from the open well were discussed in \cite{Dobrzyniecki_2018}. A recent study \cite{Brugger_2025} of the two-boson tunneling in a double-well potential indicated a strong dependence of this process on the character of the interparticle interaction: the contact interaction, although significantly oversimplifying a physical picture, qualitatively catches the main features of the phenomenon. Except for resonant transmission, another interesting effect associated with bound-state scattering that can be observed in experiments with ultracold atoms is the breakdown \cite{Amirkhanov_1966,Bondar_2010} of the left-right tunneling symmetry. The dynamics of this effect were recently studied in detail \cite{Bilokon_2025} for a few fermions in the different spin configurations. {In contrast to previous studies \cite{Gomez_2019,Olshanii_2025}, where dynamics of wave packets associated with two bosons in a bound state were studied, here we consider an asymmetric configuration with an external potential acting only on one particle.}

\section{Formulation}
\begin{figure}[h!]
	\includegraphics[width=0.475\textwidth]{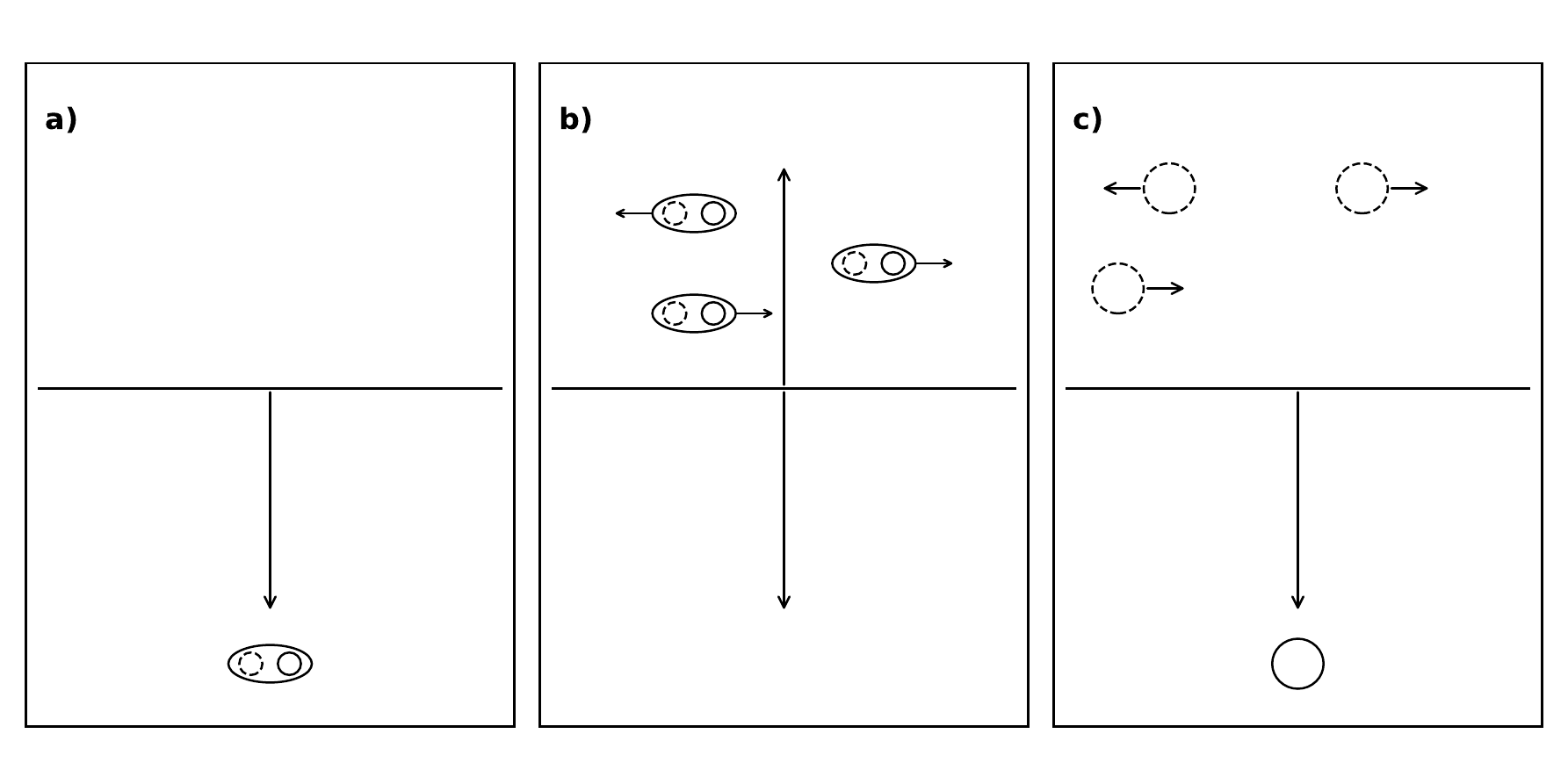}
	\caption{{Sketch of the negative-energy solutions for Hamiltonian (\ref{H}): a) corresponds to a dimer bounded by the attractive external potential; b) the dimer scattering states on both attractive and repulsive potential; c) the second particle collision with the first one localized at the origin (not discussed here). Solid circle denotes the first particle that interacts with the second one (dashed particle) and the barrier.}}\label{sketch_fig}
\end{figure}

The considered system consists of two atoms of mass $m_1$ and $m_2$, respectively, that mutually interact through the $\delta$-(pseudo)potential forming a bound state. {Both to study how the asymmetry in particle interactions with a barrier influences the dimer's scattering and to simplify further discussion, it is assumed that only one atom (say `1') scatters on the point-like defect located at the origin. Such an asymmetric interaction with an external potential is not only a theoretical construct \cite{LeBlanc_2007,Shotter_2011} but can also be realized experimentally \cite{Barker_2020} as a species-selective optical potential.} The appropriate Hamiltonian reads
\begin{eqnarray}\label{H}
	\mathcal{H}=\sum_{i=1,2}\frac{p^2_i}{2m_i}+g\delta(x_1-x_2)+g_{1}\delta(x_1),
\end{eqnarray}
in the one-dimensional case. Here $x_i$ $(i=1,2)$ are positions of atoms and $p_i$ stand for canonically conjugate momenta. {The couplings $g=-\frac{1}{ma}$ and $g_1=-\frac{1}{m_1a_1}$ can be uniquely represented via the $s$-wave scattering lengths $a$ and $a_1$.} For attractive interactions (negative $g$ and $g_1$), they can be rewritten in terms of the binding energies $\epsilon=-\frac{1}{2ma^2}$ (here $m^{-1}=m^{-1}_1+m^{-1}_2$ is the reduced mass) and $\epsilon_1=-\frac{1}{2m_1a^2_1}$. It is convenient to solve the Schr\"odinger equation {$\mathcal{H}\psi(x_1,x_2)=\mathcal{E}\psi(x_1,x_2)$} for the Hamiltonian in the momentum space, i.e., to use the Fourier harmonics representation for the wave function
\begin{eqnarray}\label{psi}
\psi(x_1,x_2)=\int_{p_1}\int_{p_2}e^{i(p_1x_1+p_2x_2)}C_{p_1,p_2},
\end{eqnarray}
(from now on $\int_{p}=\int^{\infty}_{-\infty}\frac{dp}{2\pi}$) and being interested in the bound-state solutions for {\it negative} energies $\mathcal{E}<0$, one finds [here $\varepsilon_{i,p}=\frac{p^2}{2m_i}$, $(i=1,2)$, however, the following discussion is valid for arbitrary non-quadratic one-body dispersion relations]
\begin{eqnarray}\label{C}
C_{p_1,p_2}=-\frac{c_{p_1+p_2}+c_{1,p_2}}{\varepsilon_{1,p_1}+\varepsilon_{2,p_2}-\mathcal{E}},
\end{eqnarray}
with the newly introduced functions $c_{p}$ and $c_{1,p_2}$ given by following integrals:
\begin{subequations}
\begin{align}
c_{p}=g\int dx\,e^{-ipx}\psi(x,x)=g\int_kC_{\frac{p}{2}-k,\frac{p}{2}+k}, \label{c}\\ 
c_{1,p_2}=g_1\int dx\,e^{-ip_2x}\psi(0,x)=g_1\int_{p_1}C_{p_1,p_2}. \label{c_1}
\end{align}
\end{subequations}
Note that the denominator in eq.~(\ref{C}) is always positive definite. As a result, in coordinate space, the wave function $\psi(x_1,x_2)$ tends to zero exponentially fast if $|x_1-x_2|\to \infty$, so the two particles remain bound and the dimer does not dissociate. Functions $c_{p}$ and $c_{1,k}$ of one wave-number argument each satisfy a system of coupled integral equations
\begin{subequations}
\begin{align}
t^{-1}\left(\mathcal{E}-\frac{p^2}{2M}\right)c_{p}=-\int_{k}\frac{c_{1,k}}{\varepsilon_{1,p-k}+\varepsilon_{2,k}-\mathcal{E}},\label{cc_a} \\
t_{1}^{-1}\left(\mathcal{E}-\varepsilon_{2,k}\right)c_{1,k}=-\int_{p}\frac{c_{p}}{\varepsilon_{1,p-k}+\varepsilon_{2,k}-\mathcal{E}},\label{cc_b}
\end{align}
\end{subequations}
where $M=m_1+m_2$ is the total mass of particles, and we utilized notations for the $t$-matrices $t^{-1}\left(\mathcal{E}\right)=g^{-1}+\int_{k}\frac{1}{\varepsilon_{1,k}+\varepsilon_{2,k}-\mathcal{E}}$ and $t_{1}^{-1}\left(\mathcal{E}\right)=g^{-1}_{1}+\int_{p}\frac{1}{\varepsilon_{1,p}-\mathcal{E}}$, which after evaluation of integrals, read
\begin{eqnarray}\label{tt}
t^{-1}\left(\mathcal{E}\right)=g^{-1}+\sqrt{\frac{m}{-2\mathcal{E}}}, \ \
t_{1}\left(\mathcal{E}\right)=t\left(\mathcal{E}\right)|_{g\to g_{1}, m\to m_1}.
\end{eqnarray}
The binding energies $\epsilon$ and $\epsilon_1$ are poles on the negative semi-axis of $t\left(\mathcal{E}\right)$ and $t_{1}\left(\mathcal{E}\right)$, respectively. This corresponds to negative couplings $g$ and $g_1$ in one dimension. So far, we have only assumed negative energy for the system, but to proceed with solving equations (\ref{cc_a}), (\ref{cc_b}), one needs to specify the problem more concretely. In particular, we take the sign of $g$ to be negative to provide the bound-state formation of two particles. There are three physically distinct types (see fig.~\ref{sketch_fig}) of solutions for negative $\mathcal{E}$s, {with only two of them corresponding to the dimer physics.}
The first solution, fig.~\ref{sketch_fig}(a), refers to the bound state of two particles in a potential well located at the origin. Of course, the appropriate sign of $g_1$ is required for this solution, and the wave function provides that the two particles are localized near the $x=0$ point. In this case, the energy $\mathcal{E}$ should be lower than both binding energies $\epsilon$ and $\epsilon_1$. Technically, it means that $t$-matrices $t\left(\mathcal{E}-\frac{p^2}{2M}\right)$ and $t_{1}\left(\mathcal{E}-\varepsilon_{2,k}\right)$ have no poles, and one can solve the system (\ref{cc_a}), (\ref{cc_b}) by expressing one function (say $c_{1,k}$) with the subsequent substitution into another equation. The resulting homogeneous integral equation
\begin{eqnarray}\label{c}
t^{-1}\left(\mathcal{E}-\frac{p^2}{2M}\right)c_{p}=\int_{s}\mathcal{K}_{p,s}c_{s},
\end{eqnarray}
with the symmetric kernel $\mathcal{K}_{p,s}$ given by the following integral:
\begin{eqnarray}\label{K}
\mathcal{K}_{p,s}=\int_{k}\left.\frac{t_{1}\left(\mathcal{E}\right)}{(\varepsilon_{1,p-k}-\mathcal{E})(\varepsilon_{1,s-k}-\mathcal{E})}\right|_{\mathcal{E}\to\mathcal{E}-\varepsilon_{2,k}},
\end{eqnarray}
can be solved numerically. 

The second type of solutions, fig.~\ref{sketch_fig}(b), corresponds to scattering states of the dimer formed by two particles. The appropriate energy consists of the two-body bound state energy $\epsilon$ and kinetic energy of the center-of-mass motion with momentum $P>0$. Importantly, we consider only states below the threshold $\mathcal{E}=\epsilon+\frac{P^2}{2M}<0$ for the dimer dissociation. In fact, these states had already been ignored while the solution for $C_{p_1,p_2}$ was written down. For positive total energies of the system, eq.~(\ref{C}) should be supplemented by a proper homogeneous solution, and to avoid a pole, one must shift the energy in the upper complex half-plane $\mathcal{E}\to\mathcal{E}+i0_+$. Even below the threshold $\mathcal{E}=0$, the bound state scattering on potential located at the origin suggests that $t^{-1}\left(\mathcal{E}-\frac{p^2}{2M}\right)$ (with $\mathcal{E}=\epsilon+\frac{P^2}{2M}$) crosses zero at $|p|=P$. Therefore, equating a function $c_{p}$ from eq.~(\ref{c}), one is free to add in the right-hand side term proportional to the $\delta$-function $\delta(|p|-P)$. The boundary conditions dictate the correct form for this homogeneous solution. In a case of the dimer scattering, the wave function at large distances from the $x=0$ point is a product of the localized two-body bound state wave function that depends on the relative coordinate, and the plane wave for the center-of-mass motion. This statement is true only in dimensions higher than one. The generic scattering state is asymptotically given by the plane wave plus an outgoing spherical wave from the origin. In the one-dimensional case, the spherical wave does not vanish at infinity giving the well-known asymptotic forms of the wave function before $\psi(x_1,x_2)|_{x_i\to -\infty}=\left[e^{iP(m_1x_1+m_2x_2)/M}-r_Pe^{-iP(m_1x_1+m_2x_2)/M}\right]\phi(|x_1-x_2|)$ and after $\psi(x_1,x_2)|_{x_i\to \infty}=t_Pe^{iP(m_1x_1+m_2x_2)/M}\phi(|x_1-x_2|)$ the scattering center (here $\phi(|x_1-x_2|)$ is the two-body bound-state wave function when $g_1=0$). Matching the limiting behavior of the scattering wave function at infinity to the general solution $\ref{C}$ leads to the following choice (recall, $\mathcal{E}=\epsilon+\frac{P^2}{2M}$):
\begin{eqnarray}\label{c_scatt}
c_{p}=2\pi\delta(p-P)
+t\left(\mathcal{E}+i0_+-\frac{p^2}{2M}\right)\int_{s}\mathcal{K}_{p,s}c_{s},
\end{eqnarray}
the `$i0_+$-prescription' should also be applied to the energy dependence in kernel $\mathcal{K}_{p,s}$. Now one can explicitly verify by putting the obtained solution for $c_{p}$ into eq.~(\ref{C}) and then in eq.~(\ref{psi}) that the first $\delta$-function term guarantees a proper behavior at $|x_1|,|x_2|\to \infty$. In contrast, the impact of the second term and function $c_{1,p_2}$ is negligible in this limit.

The third type, fig.~\ref{sketch_fig}(c), of negative-energy solutions for the system of equations (\ref{cc_a}), (\ref{cc_b}) not discussed below, is the scattering states of the second particle when the first one is in the bound state localized by the potential at the origin. The latter case, of course, suggests the `correct' sign of the coupling $g_1$. The solution can be constructed similarly, in terms of a linear non-homogeneous integral equation with a different kernel, but now for the function $c_{1,k}$.

By construction, eq.~(\ref{c_scatt}) is similar to the scattering solution for the particle moving in the potential, an even function of $x$. This is not a coincidence because the function $c_{p}$ depends only on the total momentum of the two-body bound state and effectively solves the Schr\"odinger equation for the center-of-mass motion. Furthermore, at very small $P$s, when the de Broglie wavelength is large in comparison to $a$ and $a_1$, the solution of (\ref{c_scatt}) should asymptotically recover the wave function of a particle with mass $M$ that moves in the $\delta$-potential. The latter observation prompts the ansatz:
\begin{eqnarray}\label{c_ansatz}
c_{p}=2\pi\delta(p-P)+
\frac{f_{p}(\mathcal{E})}{p^2-P^2-i0_+},
\end{eqnarray}
where the introduced function $f_{p}(\mathcal{E})$ at $|p|=P$ (on-shell), reproduces the two-particle bound-state scattering amplitude. The latter determines the wave function at large distances from the scattering center. Plugging the ansatz into eq.~(\ref{c_scatt}), one obtains an equation for the off-shell amplitude
\begin{eqnarray}\label{f_p}
f_{p}(\mathcal{E})=h_{p}(\mathcal{E})\left[\mathcal{K}_{p,P}+\int_{s}\frac{\mathcal{K}_{p,s}f_{s}(\mathcal{E})}{s^2-P^2-i0_+}\right],
\end{eqnarray}
with the introduced even function $h_{p}(\mathcal{E})=(p^2-P^2)t\left(\mathcal{E}+i0_+-\frac{p^2}{2M}\right)$ of the wave number $p$. In higher dimensions, the analog of eq.~(\ref{f_p}) allows the spherical harmonics expansion due to conservation of the two-particle angular momentum. This is an infinite series over the Gegenbauer polynomials \cite{Abramowitz}. Their form strongly depends on the spatial dimension. The one-dimensional integral equation (\ref{f_p}) preserves parity and consequently holds the discrete symmetry, suggesting the following exact decomposition: $f_{p}(\mathcal{E})=f^{+}_{|p|}(\mathcal{E})+\textrm{sign}(p)f^{-}_{|p|}(\mathcal{E})$. Functions $f^{+}_{|p|}(\mathcal{E})$ and $f^{-}_{|p|}(\mathcal{E})$ are an analog of the $s$-wave and $p$-wave partial scattering amplitudes that {\it exactly} determine the solution in the one-dimensional case, and by utilizing a similar decomposition for the kernel $\mathcal{K}_{p,s}=\mathcal{K}^{+}_{|p|,|s|}+\textrm{sign}(ps)\mathcal{K}^{-}_{|p|,|s|}$, it is easy to obtain a set of independent integral equations for them.

\section{Results and discussion}
\begin{figure}[h!]
	\centerline{\includegraphics
		[width=0.45
		\textwidth,clip,angle=-0]{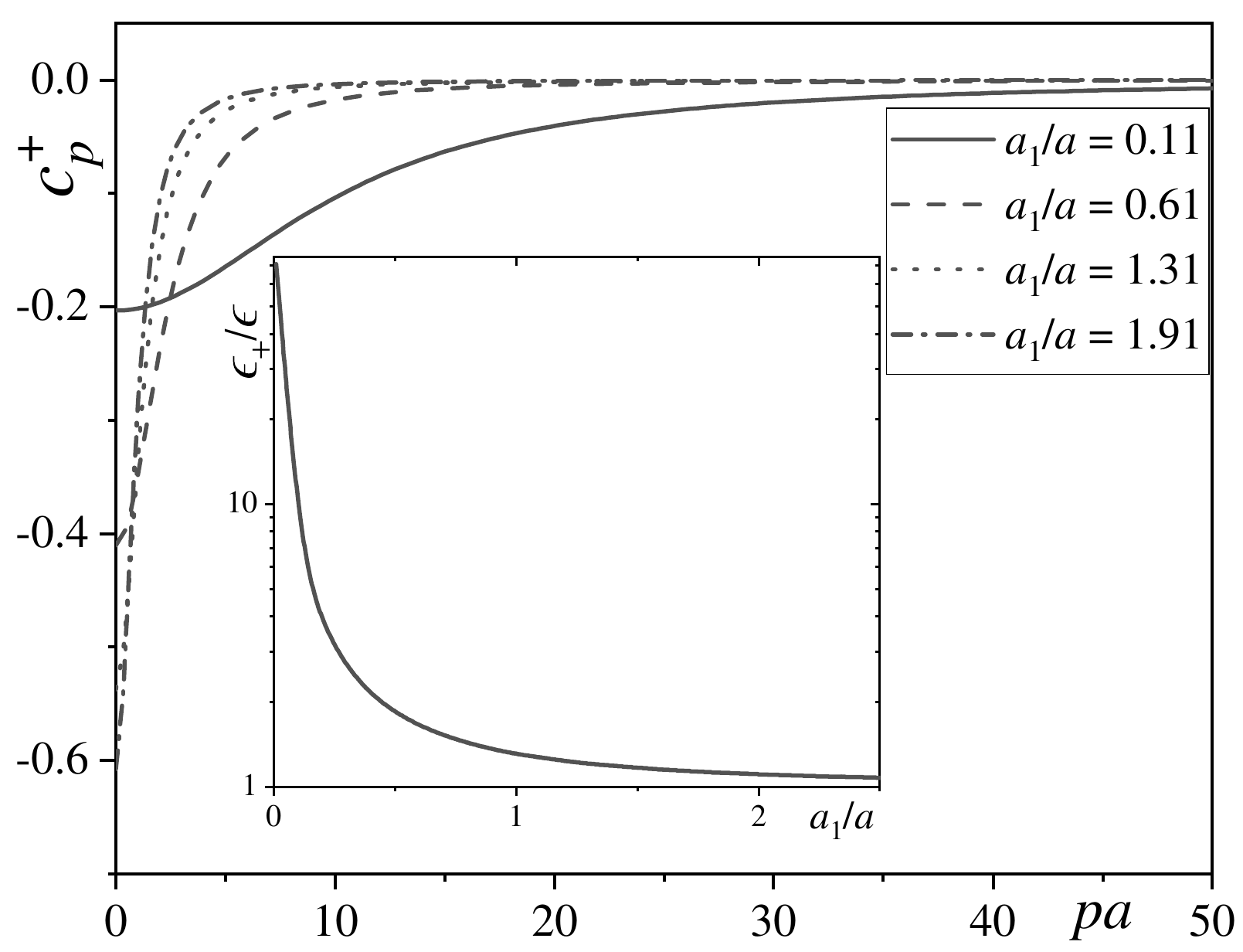}}
	\caption{Symmetric non-normalized auxiliary function  $c^{+}_p$ that determines the dimer bound states in external potential at different couplings. The insert shows (in logarithmic scale) the appropriate energy $\epsilon_+$ in units of $\epsilon=-\frac{1}{2ma^2}$.}\label{bound_state_1d_fig}
\end{figure}

To simplify further discussion, we assume that although non-identical, the two particles have the same masses $ m_1 = m_2$. Properties of the system at non-unit mass ratios are not qualitatively different. Coupling $g$ should be negative to support the two-particle bound state, and there are no restrictions on the sign of $g_1$ for the dimer scattering. For the states of dimer localized at the origin, i.e., described by equation (\ref{c}) $g_1<0$ is required. Results for the dimer's symmetric bound states obtained by numerical diagonalization of eq.~\ref{c} are presented in fig.~\ref{bound_state_1d_fig}. 
The numerical procedure was implemented by transforming an integral equation into a matrix equation. Depicted {auxiliary functions $c^{+}_p$ that determine the bound state wave functions with eigenstates $\epsilon_+$ clearly} demonstrate that this is the ground state of the system. {Indeed, since $c^{+}_p$ is real, even and sign-definite function, and $t_{1}^{-1}\left(\mathcal{E}-\varepsilon_{2,k}\right)<0$ (for any $k$ and energies $\mathcal{E}$ below threshold), signs of $c_{1,k}$ and $c^{+}_p$ coincide, providing the wave function in momentum space, $C_{p_1,p_2}$, is real and has no nodes.} The behavior of energy $\epsilon_+$ (see insert) in two limiting cases can be understood by simple arguments. At weak dimer couplings $a_1/a\ll 1$, the ground-state energy, up to the perturbative corrections, is given by the binding energy of the first atom in the $\delta$-potential $\epsilon_+/\epsilon\approx \frac{m_2a^2}{Ma^2_1}$. In the inverse limit $a/a_1\ll 1$, the dimer can be thought of as a point-like particle of mass $M$ that interacts with the potential well located at the origin, and consequently we have $\epsilon_+/\epsilon\approx 1+ \frac{m_2a^2}{m_1a^2_1}$. {We have not found excited even-parity bound states, as well as antisymmetric states with energies $\epsilon_{-}$ less than the dimer binding energy $\epsilon$, at least in the equal-mass limit.} 

\begin{figure}[h!]
	\includegraphics[width=0.235\textwidth]{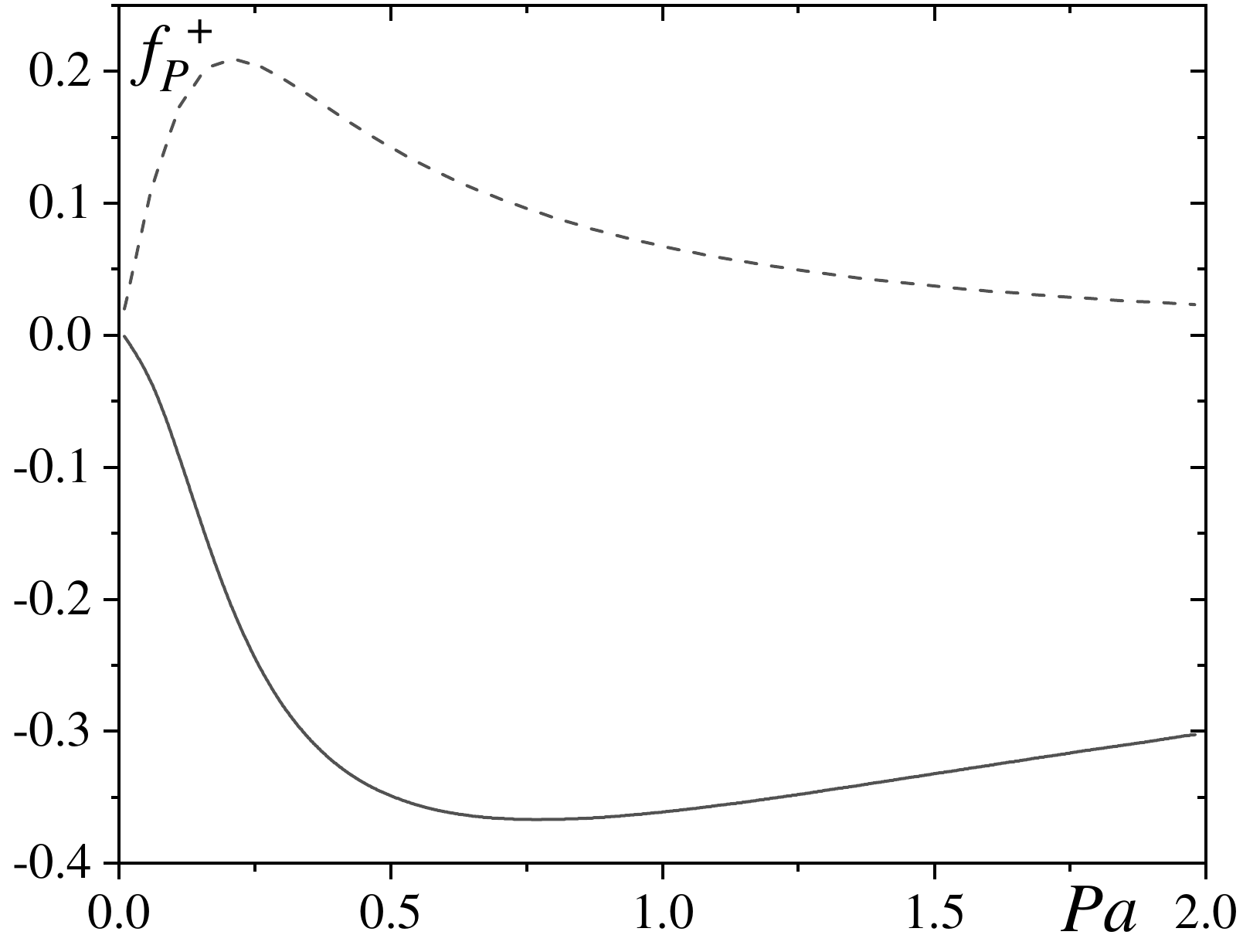}
	\includegraphics[width=0.235\textwidth]{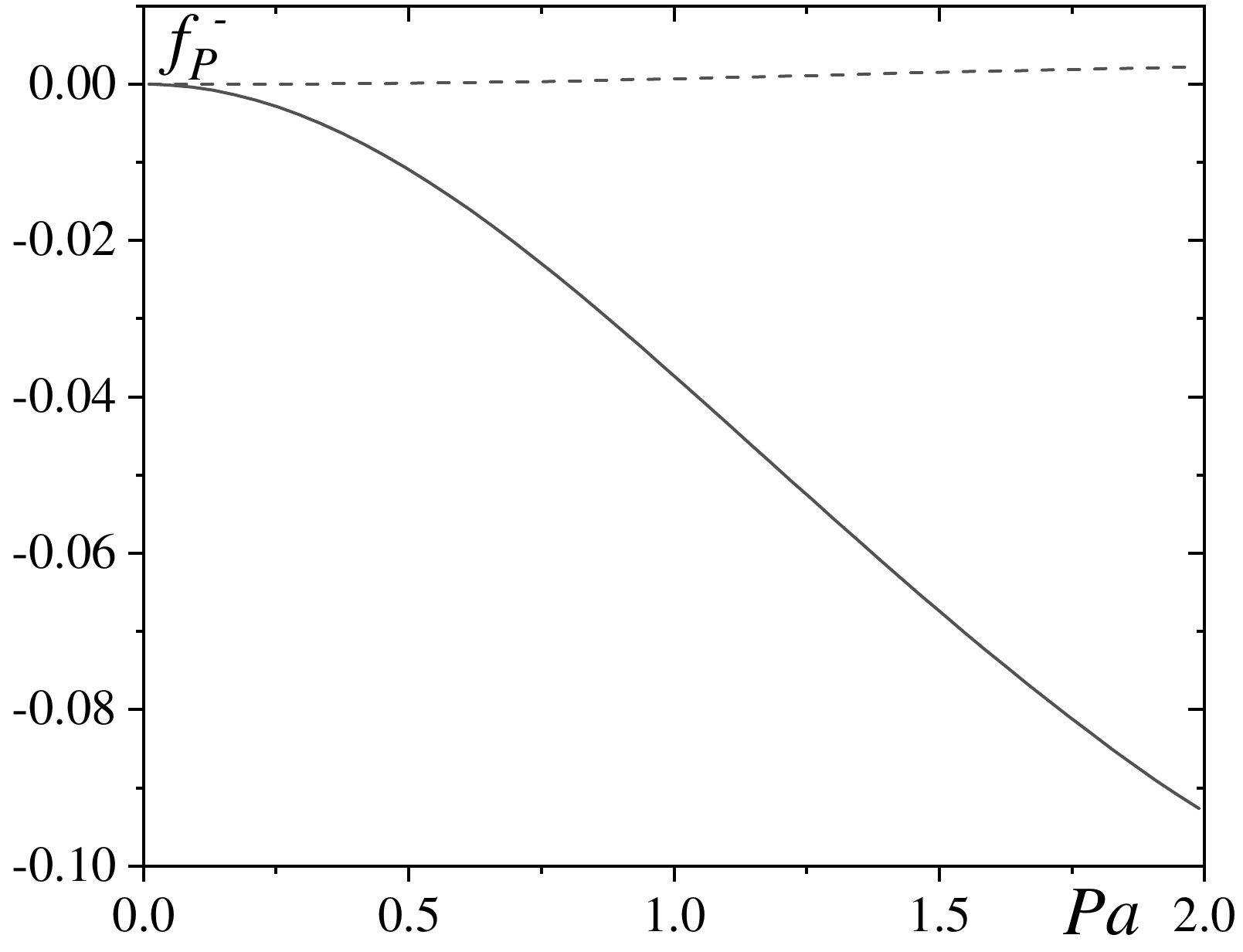}
	\includegraphics[width=0.235\textwidth]{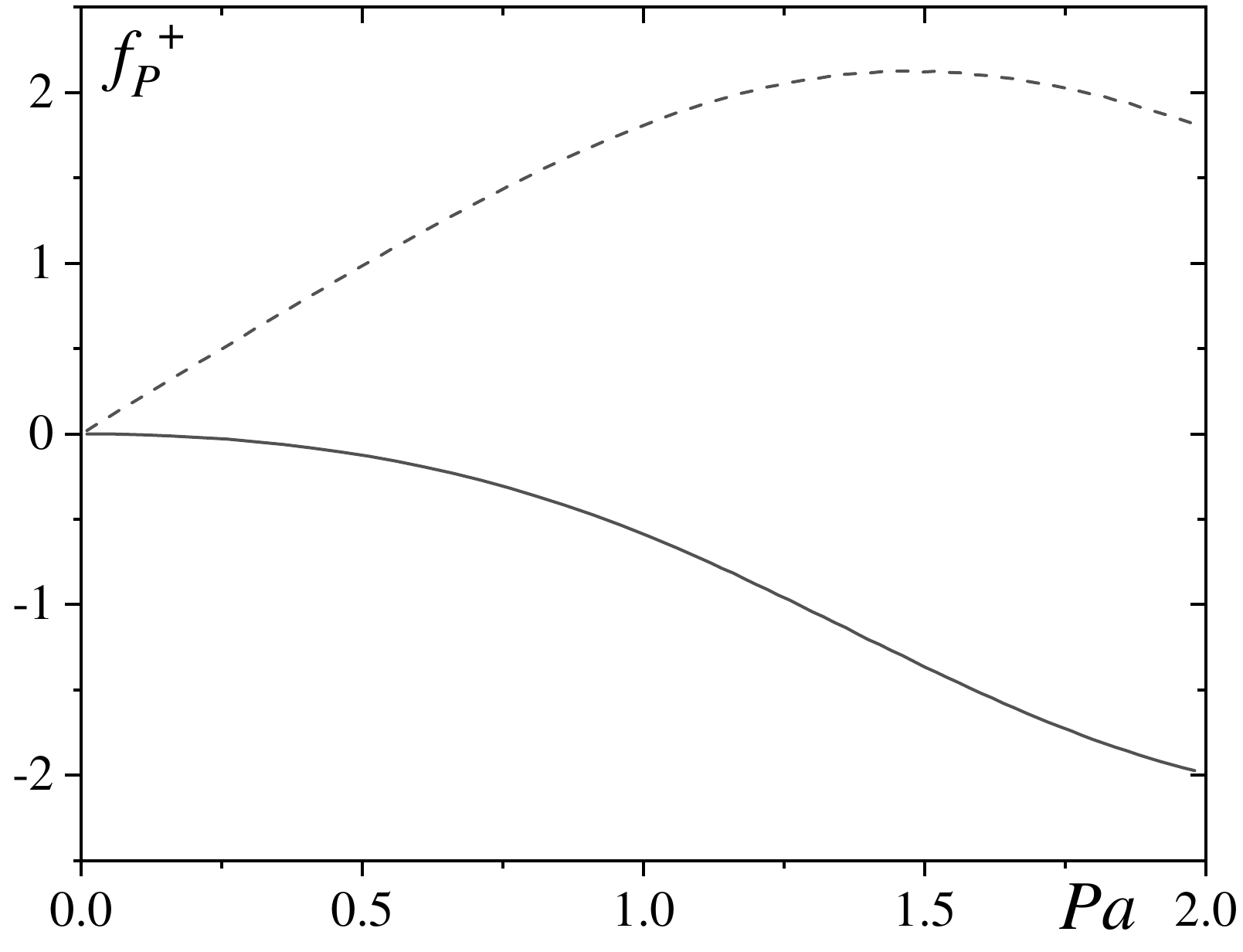}
	\includegraphics[width=0.235\textwidth]{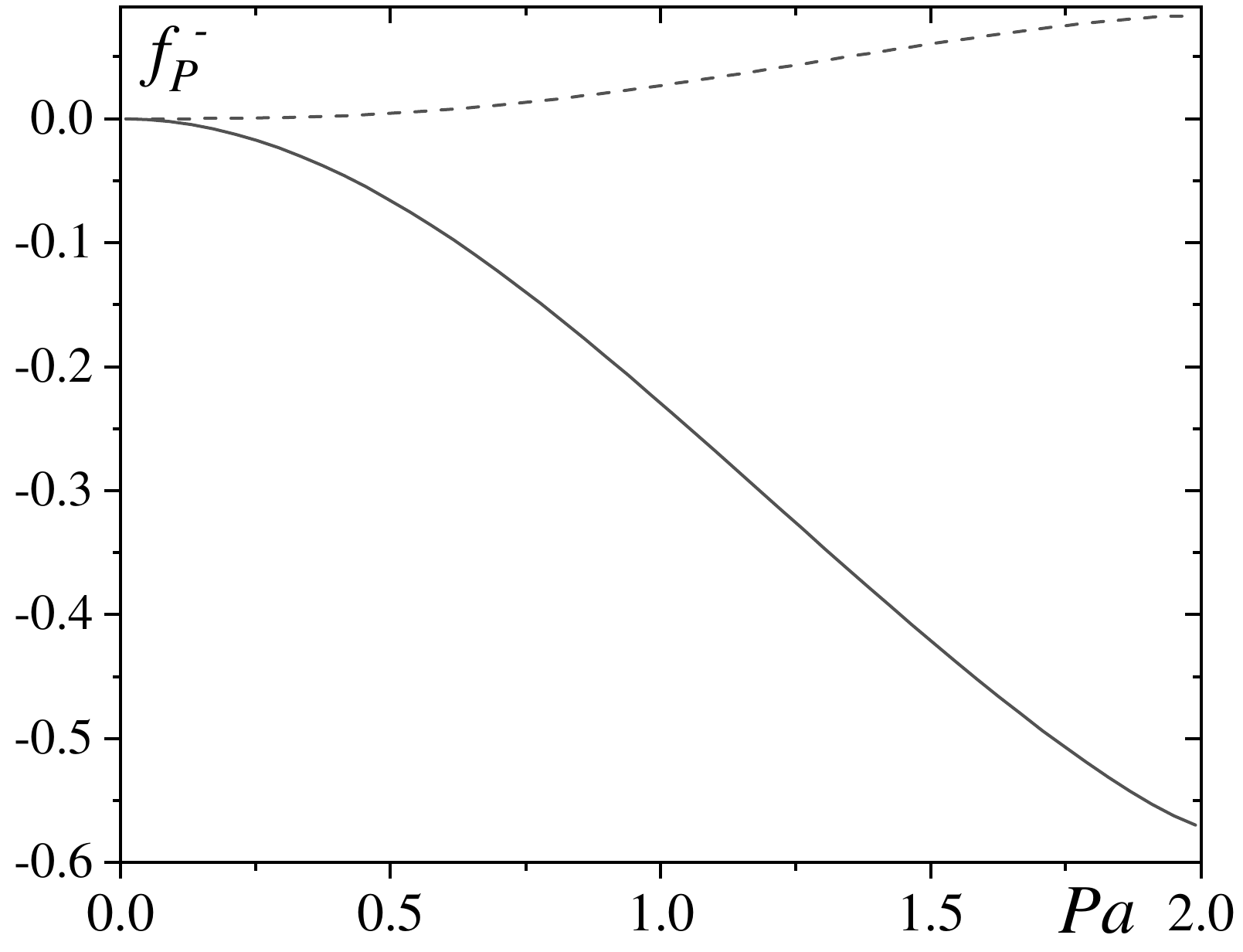}
	\includegraphics[width=0.235\textwidth]{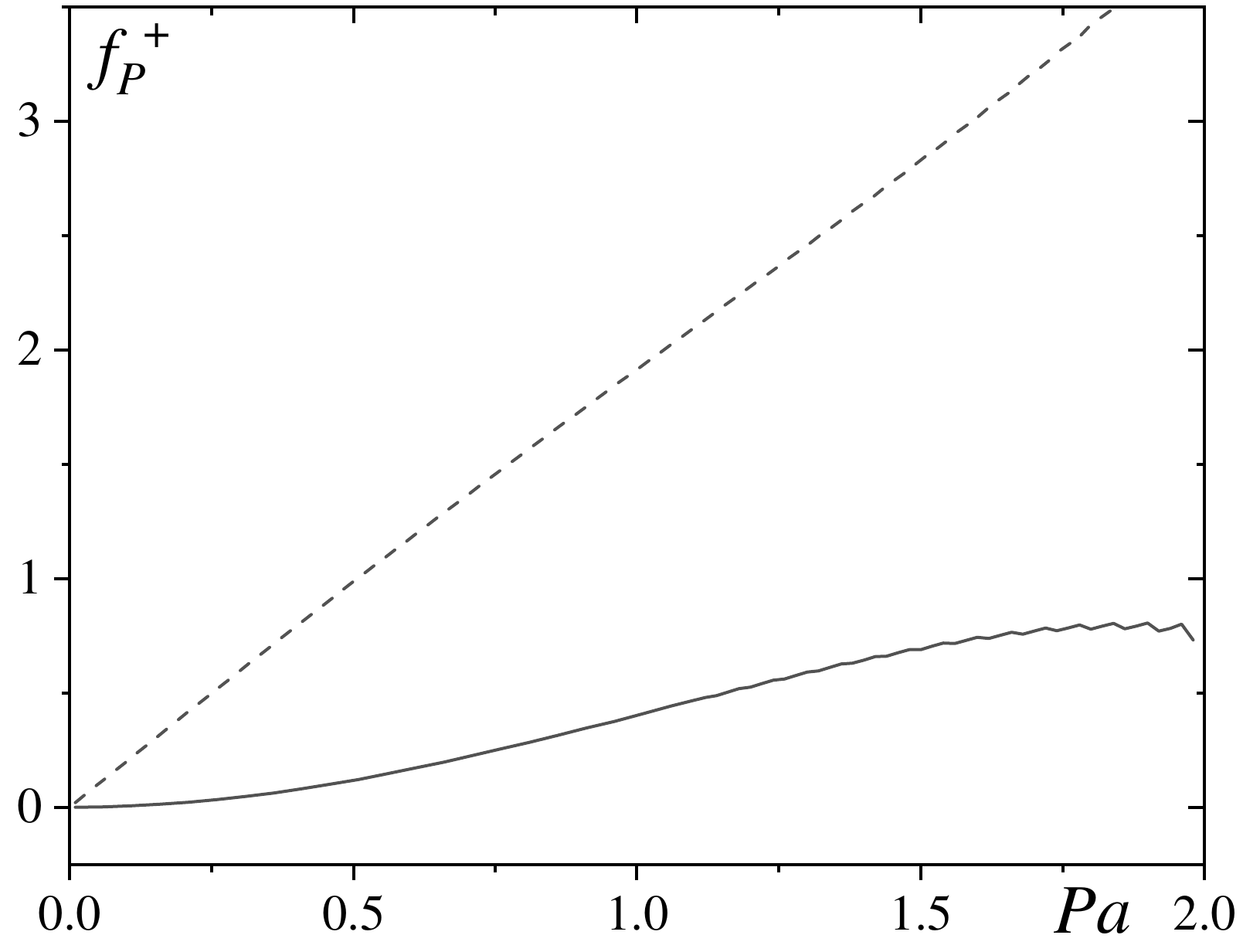}
	\includegraphics[width=0.235\textwidth]{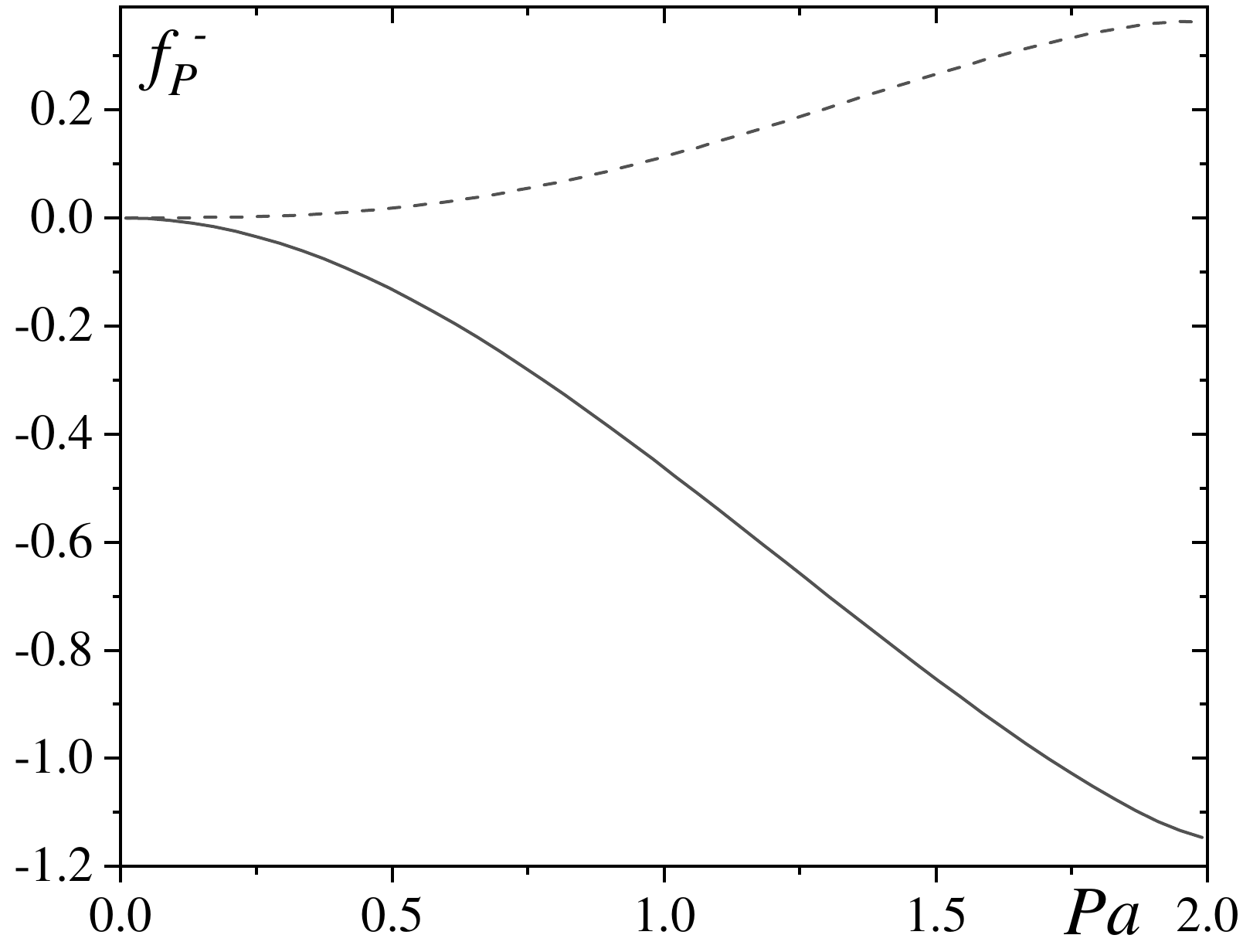}
	\caption{Symmetric (left) and antisymmetric (right) on-shell scattering amplitudes {(in units of $1/a$)} for negative $a_1$ (repulsive potential) from weak (top) to strong (bottom) couplings $a_1/a=-10; \ -1; \ -0.1$. {Solid and dashed lines correspond to the real and imaginary parts of the scattering amplitude, respectively.}}\label{1d_rep_scatt_fig}
\end{figure}

A similar numerical procedure was employed to calculate the on-shell symmetric and antisymmetric scattering amplitudes $f^{\pm}_{P}\left(\epsilon+\frac{P^2}{2M}\right)$, {both measured in units of $1/a$}. The obtained curves that are presented in fig.~\ref{1d_rep_scatt_fig}
demonstrate the character of the dimer interaction with the repulsive ($g_1>0$) scattering center: although $f^{-}_{P}\left(\epsilon+\frac{P^2}{2M}\right)$ qualitatively maintains its behavior at strong coupling, the symmetric scattering amplitude changes sign of the real part signaling the shift of the dimer-barrier effective interaction from repulsion to attraction. It can be tracked from the low-energy expansion of $f^{+}_P\left(\epsilon+\frac{P^2}{2M}\right)\asymp \frac{2iP}{1+iPa_d}$ (here $a_d$ is the dimer-barrier scattering length). Note that $Pa=2$ is a threshold for the dimer to dissociate into two equal-mass particles. {This is the boundary, where the dimer's total energy $\mathcal{E}=0$.} At positive $a_1$, the threshold occurs at lower momenta and for small ratios $a_1/a\le 1/\sqrt{2}$ it is energetically more preferable for the first particle to be trapped by the $\delta$-well even at $P=0$. For larger ratios, there is always a maximal center-of-mass momentum $Pa/2=\sqrt{1-a^2/2a^2_1}$ {(this is just a reformulated condition $\epsilon+\frac{P^2}{2M}= \epsilon_1$ in the equal-mass $m_1=m_2$ limit)} above which our analysis fails. 
\begin{figure}[h!]
	\includegraphics[width=0.235\textwidth]{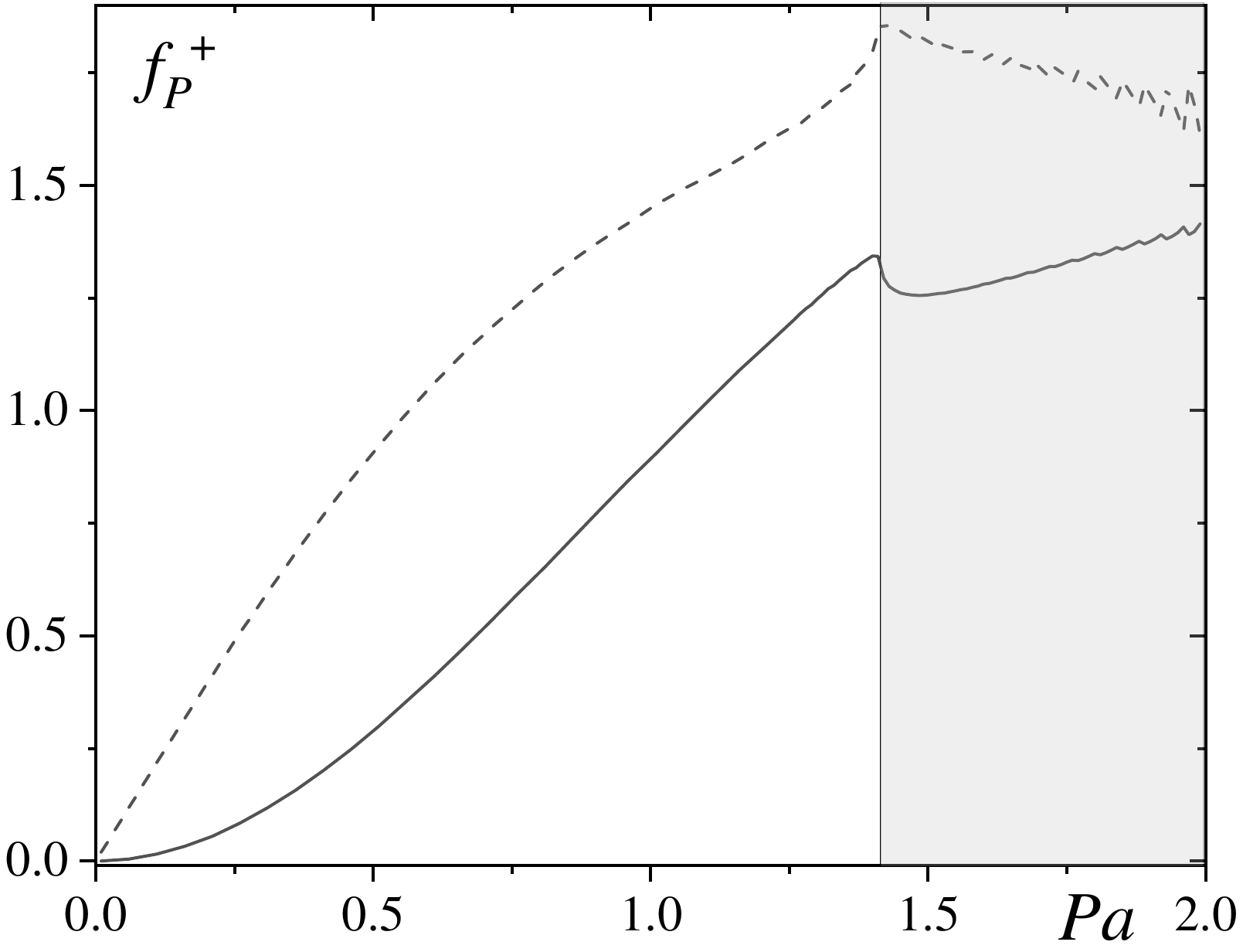}
	\includegraphics[width=0.235\textwidth]{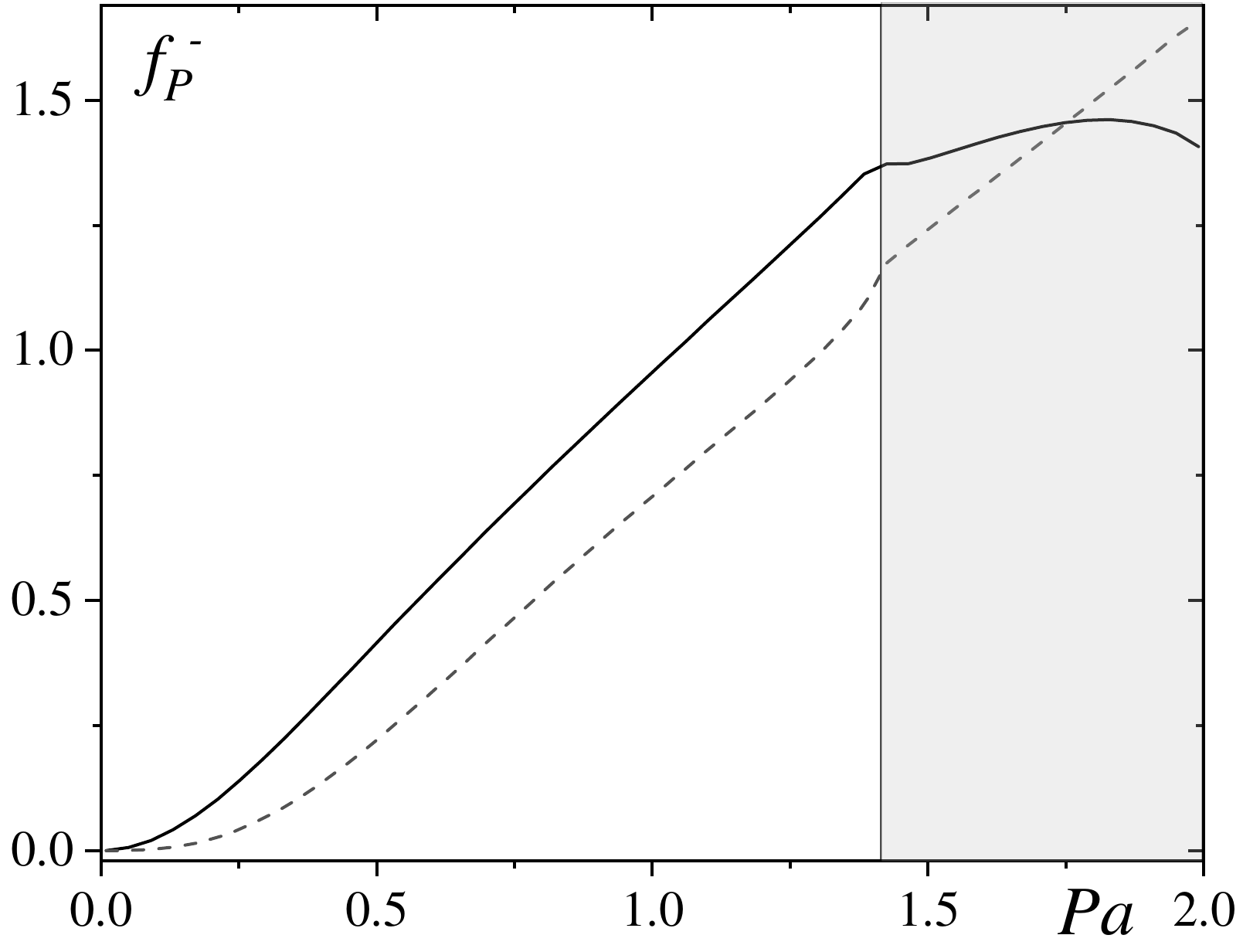}
	\includegraphics[width=0.235\textwidth]{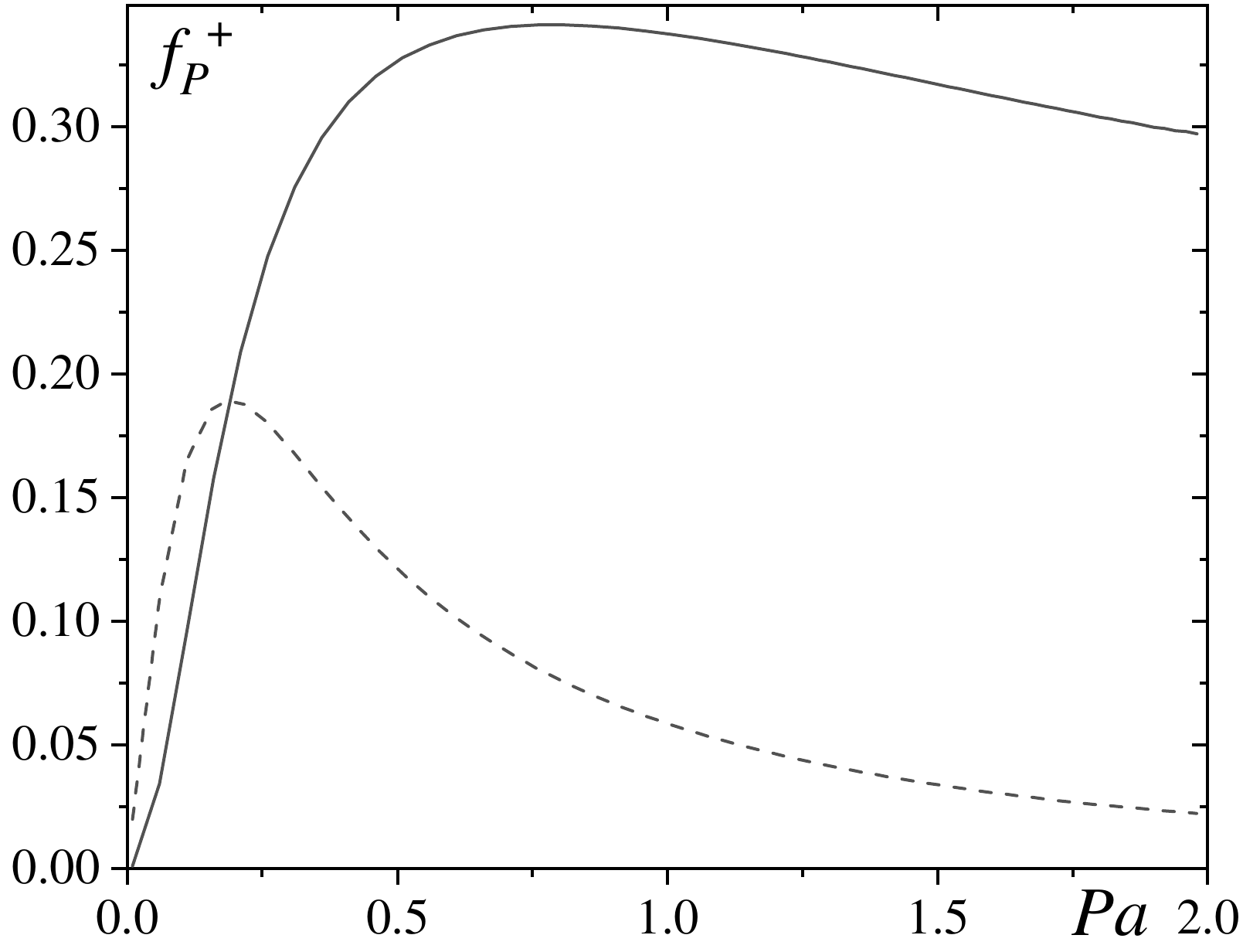}
	\includegraphics[width=0.235\textwidth]{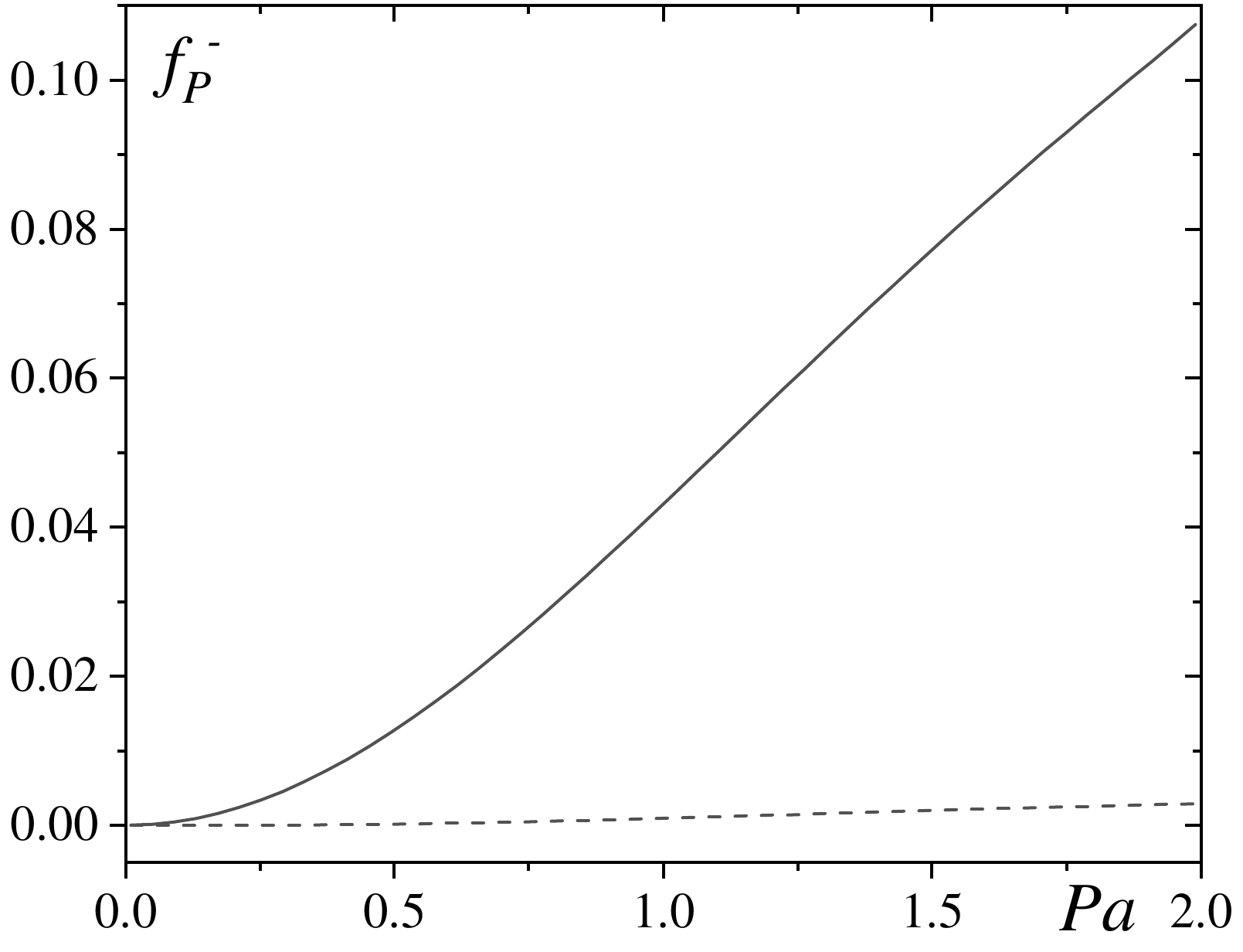}
	\caption{Even (left) and odd (right) on-shell scattering amplitudes {(in units of $1/a$)} for attractive interaction (positive $a_1$) $a_1/a=1$ (top) and $a_1/a=10$ (bottom). Solid and dashed lines correspond to the real and imaginary parts of the scattering amplitude, respectively. Shaded areas represent the regions ($Pa\ge \sqrt{2}$ for $a_1/a=1$,  and $Pa\ge 1.995$ not shown for $a_1/a=10$) of inapplicability of the considered scattering solution (\ref{c_scatt}).}\label{1d_att_scatt_fig}
\end{figure}
Figure~\ref{1d_att_scatt_fig} presents the on-shell scattering amplitudes for two attractive couplings. The shaded area denotes the region in parameter space, where the wave function (\ref{c_scatt}) is incomplete due to the disregard of states with the dissociated dimer.

\begin{figure}[h!]
	\includegraphics[width=0.45\textwidth]{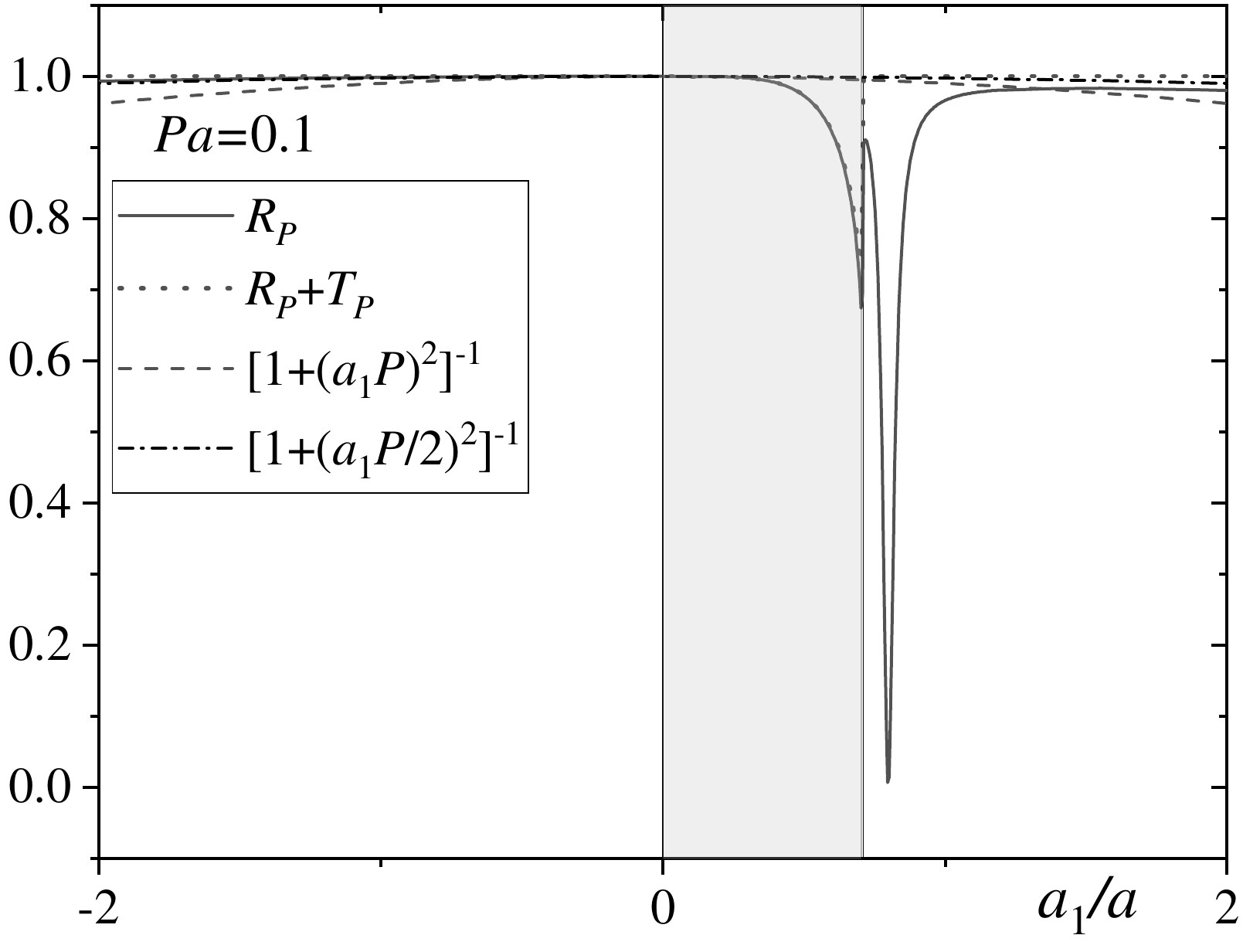}
	\includegraphics[width=0.45\textwidth]{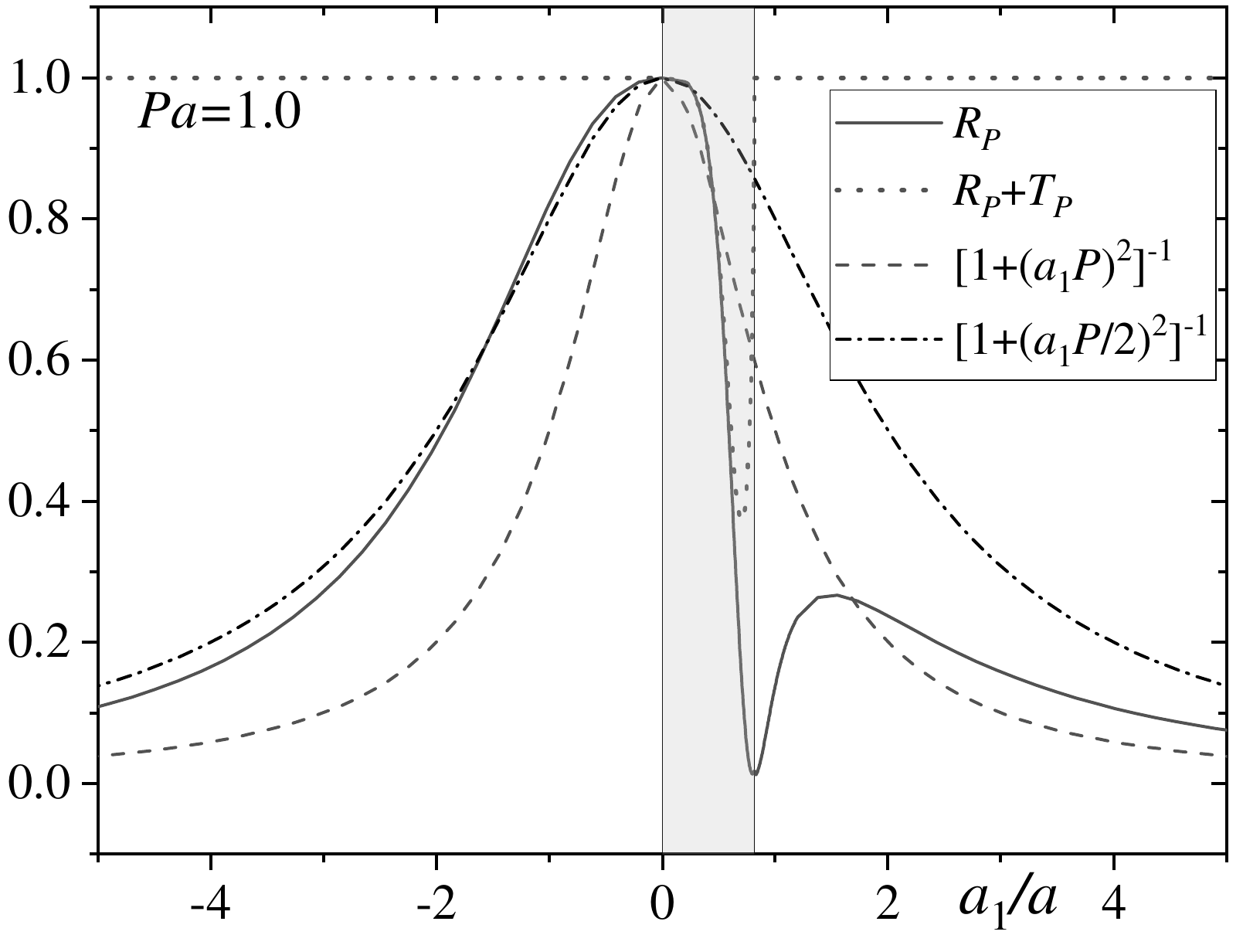}
	\caption{The dimer reflection probabilities as a function of the potential barrier (well) height (depth) for two center-of-mass momenta $Pa=0.1$ (top) and $Pa=1.0$ (bottom). Small $a_1/a$ corresponds to strong repulsion (negative ratios) and strong attraction (positive ratios). In the shaded areas, the dimer can dissociate into two particles, which is not accounted for in the present analysis.}\label{R_P_fig}
\end{figure}

Other observable quantities can be extracted from the on-shell scattering amplitude, namely the reflection $R_P$ and transmission $T_P$ probabilities
\begin{eqnarray}\label{RT_P}
R_P=\left|\frac{f_{-P}(\mathcal{E})}{2P}\right|^2,\ \ T_P=\left|1-\frac{f_{P}(\mathcal{E})}{2iP}\right|^2,
\end{eqnarray}
with $\mathcal{E}=\epsilon+\frac{P^2}{2M}$. These expressions are straightforwardly derived from the asymptotic behavior of the dimer scattering wave function $\psi(x_1,x_2)$ at large center-of-mass coordinates. We have calculated both quantities (see fig.~\ref{R_P_fig}) to verify the accuracy of our numerical procedure by the fulfillment of the equality $T_P+R_P=1$. In the region of parameter space where the solution (\ref{c_scatt}) is complete, the maximal deviation from unity was found to be $10^{-5}$. Since there is only a single inelastic channel in the shaded region, the deviation $1-T_P-R_P$ can be treated as a probability for dimer dissociation. To reveal the dimer scattering properties, we compared the numerically calculated $R_P$
to the $a_1$-dependence of the reflection probability $\frac{1}{1+(Pa_1)^2}$ for a single atom `1' at two fixed momenta $Pa=0.1$ and $Pa=1.0$. Since the smallness of the dimer kinetic energy in the former case, the shaded region is almost restricted to $a_1/a\le 1/\sqrt{2}$. For $Pa=1.0$, the region stretches to $a_1/a\le \sqrt{2/3}$ (where $\epsilon+\frac{P^2}{2M}\ge \epsilon_1$). Ignoring regions of inapplicability of the results obtained, a generic tendency is clearly seen: it is harder for the dimer to pass through the potential barrier. Even in the limit when the dimer size $a$ is negligible compared to the barrier height $a_1$, the dimer reflection coefficient can be approximated by the Lorentzian $\frac{1}{1+(Pa'_1)^2}$ with some effective scattering length $a'_1/a_1=m_1/M<1$ independently of the coupling $g_1$ sign. The latter circumstance immediately leads to the enhancement of the reflection probability. At a strong enough attractive potential, however, we find regions $0.7\lesssim a_1/a\lesssim 1.2$ for $Pa=0.1$ and $0.8\lesssim a_1/a\lesssim 1.8$ for $Pa=1.0$, where the transmission coefficient can be substantially larger than its single-particle counterpart.

\section{Summary}
In conclusion, we have presented a detailed study of the quantum-mechanical one-dimensional problem of two coupled non-identical particles that form a dimer when one of them interacts with the external contact potential. By solving the Schr\"odinger equation in momentum representation, we identified three types of solutions, focusing on the two that do not suggest the dimer dissociation. These are: the localized states when the dimer as a whole is trapped by the attractive external potential, and scattering states with the total energy of the system larger than the two-particle binding energy. In the equal-mass limit of two particles, we identified only {even} localized states and numerically calculated the even- and odd-parity partial contributions to the on-shell scattering amplitude. Additionally, we analyzed the dependence of the dimer reflection coefficient as a function of the potential barrier height at fixed center-of-mass momentum. In general, the dimer reflection is enhanced in comparison to a single-particle counterpart; however, for an attractive potential, there is a narrow region in parameter space of significant growth of the dimer transmission probability.

Extending our discussion to higher dimensions and a larger number of $\delta$-potentials \cite{Panochko_2021} would be interesting. It is also essential to determine whether the described resonant transmission region survives in quasi-one-dimensional and quasi-two-dimensional geometries \cite{Hryhorchak_2023_2}. Another possible direction is to study a system with two particles interacting with the external potential. Cases of two non-identical particles and two bosons can reveal the role of statistical effects in dimer scattering.

\begin{center}
		{\bf Acknowledgements}
\end{center}      
We are grateful to all colleagues from our department for discussing the results. In the final stage, O.H.'s work was supported by Project 2025.07/0326 (No.~0126U002943) from the National Research Foundation of Ukraine.


\begin{thebibliography}{99}

\bibitem{Zakhariev_1964} B. Zakhariev and S. Sokolov, {\it Intensified Tunnel Effect for Complex Particles.} \href{https://doi.org/10.1002/andp.19644690502}{Annalen der Physik {\bf 469}, 229 (1964).} 

\bibitem{Saito_1994} N. Saito and Y. Kayanuma, {\it Resonant tunnelling of a composite particle through a single potential barrier.} \href{https://doi.org/10.1088/0953-8984/6/20/014}{J. Phys.: Condens. Matter {\bf 6}, 3759 (1994).}

\bibitem{Bertulani_2015} C. A. Bertulani, {\it Tunneling of Atoms, Nuclei and Molecules.} \href{https://doi.org/10.1007/s00601-015-0990-z}{Few-Body Syst. {\bf 56}, 727 (2015).}
\bibitem{Razavy} M. Razavy, {\it Quantum theory of tunneling}, 2nd ed.
(World Scientific, Singapore 2013).

\bibitem{Bertulani_2007} C. A. Bertulani, V. V. Flambaum and V. G. Zelevinsky, {\it Tunneling of a composite particle: effects of intrinsic structure.} \href{https://doi.org/10.1088/0954-3899/34/11/006}{J. Phys. G: Nucl. Part. Phys. {\bf 34}, 2289 (2007).}

\bibitem{Krassovitskiy_2014} P. M. Krassovitskiy and F. M. Pen'kov, {\it Contribution of resonance tunneling of molecule to physical observables.} \href{https://doi.org/10.1088/0953-4075/47/22/225210}{J. Phys. B: At. Mol. Opt. Phys. {\bf 47}, 225210 (2014).}

\bibitem{Bacca_2006} S. Bacca and H. Feldmeier, {\it Resonant tunneling in a schematic model.} \href{https://doi.org/10.1103/PhysRevC.73.054608}{Phys. Rev. C {\bf 73}, 054608  (2006).}

\bibitem{Takigawa_1984} N. Takigawa and G. F. Bertsch, {\it Semiclassical theory of quantum tunneling in multidimensional systems.} \href{https://doi.org/10.1103/PhysRevC.29.2358}{Phys. Rev. C {\bf 29}, 2358 (1984).} 

\bibitem{Penkov_2000} F. M. Pen'kov, {\it Metastable states of a coupled pair on a repulsive barrier.} \href{https://doi.org/10.1103/PhysRevA.62.044701}{Phys. Rev. A {\bf 62}, 044701 (2000).}

\bibitem{Sato_2002} T. Sato and Y. Kayanuma, {\it Quantum inelasticity in reflection of a composite particle.} \href{https://doi.org/10.1209/epl/i2002-00246-4}{Europhys. Lett. (EPL) {\bf 60}, 331 (2002).}

\bibitem{Goodvin_2005} G. L. Goodvin and M. R. A. Shegelski, {\it Tunneling of a diatomic molecule incident upon a potential barrier.} \href{https://doi.org/10.1103/PhysRevA.71.032719}{Phys. Rev. A {\bf 71}, 032719 (2005).}

\bibitem{Ahsan_2010} N. Ahsan and A. Volya, {\it Quantum tunneling and scattering of a composite object reexamined.} \href{https://doi.org/10.1103/PhysRevC.82.064607}{Phys. Rev. C {\bf 82}, 064607 (2010).}

\bibitem{Lee_2006} Y. J. Lee, {Journal of the Korean Physical Society {\bf 49}, 103 (2006).}

\bibitem{Hnybida_2008} J. Hnybida and M. R. A. Shegelski, {\it Tunneling of a diatomic molecule with many bound states.} \href{https://doi.org/10.1103/PhysRevA.78.032711}{Phys. Rev. A {\bf 78}, 032711 (2008).}

\bibitem{Gusev_2014} Gusev, A.A. et al., {\it Resonant tunneling of a few-body cluster through repulsive barriers.} \href{https://doi.org/10.1134/S1063778814030107}{Phys. Atom. Nuclei {\bf 77}, 389 (2014).}

\bibitem{Shegelski_2019} M. R. A. Shegelski,  et al., {\it Resonant transmission of weakly bound multi-atomic molecules.} \href{https://doi.org/10.1088/1361-6455/aafa4a}{J. Phys. B: At. Mol. Opt. Phys. {\bf 52}, 055201 (2019).}

\bibitem{Shegelski_2008} M. R. A. Shegelski, J. Hnybida, H. Friesen, C. Lind, and J. Kavka, {\it Tunneling of a diatomic molecule with unbound states in one dimension.} \href{https://doi.org/10.1103/PhysRevA.77.032702}{Phys. Rev. A {\bf 77}, 032702 (2008).}

\bibitem{Shegelski_2008_2} M.  R. A. Shegelski, J. Hnybida, and R. Vogt, {\it Formation of a molecule by atoms incident upon an external potential.} \href{https://doi.org/10.1103/PhysRevA.78.062703}{Phys. Rev. A {\bf 78}, 062703 (2008).}






\bibitem{Hryhorchak_2023} O. Hryhorchak and V. Pastukhov, {\it Second root of dilute Bose-Fermi mixtures.} \href{https://doi.org/10.1088/1751-8121/accda4}{J. Phys. A: Math. Theor. {\bf 56}, 205003 (2023).}

\bibitem{Lee_2011} D. Lee and M. Pine, {\it How quantum bound states bounce and the structure it reveals.} \href{https://doi.org/10.1140/epja/i2011-11041-4}{Eur. Phys. J. A {\bf 47}, 41 (2011).}
\bibitem{Pine_2013} M. Pine and D. Lee, {\it Effective field theory for bound state reflection.} \href{https://doi.org/10.1016/j.aop.2012.12.009}{Ann. Phys. {\bf 331}, 24 (2013).}


\bibitem{Sowinski_2019} T. Sowi\ifmmode \acute{n}\else \'{n}\fi{}ski and M. \'A. Garc\'{\i}a-March, {\it One-dimensional mixtures of several ultracold atoms: a review.} \href{https://doi.org/10.1088/1361-6633/ab3a80}{Rep. Prog. Phys. {\bf 82}, 104401 (2019).} 
\bibitem{Mistakidis_2023} S. I. Mistakidis, A. G. Volosniev, R. E. Barfknecht, T. Fogarty, T. Busch, A. Foerster, P. Schmelcher, and
N. T. Zinner, {\it Few-body Bose gases in low dimensions---A laboratory for quantum dynamics.} \href{https://doi.org/10.1016/j.physrep.2023.10.004}{Phys. Rep. {\bf 1042}, 1 (2023).}


\bibitem{Zurn_2012} G. Z\"urn, F. Serwane, T. Lompe, A. N. Wenz, M. G. Ries, J. E. Bohm, S. Jochim, {\it Fermionization of Two Distinguishable Fermions.} \href{https://doi.org/10.1103/PhysRevLett.108.075303}{Phys. Rev. Lett. {\bf 108}, 075303 (2012).}

\bibitem{Zurn_2013} G. Z\"urn, A. N. Wenz, S. Murmann, A. Bergschneider, T. Lompe, S. Jochim, {\it Pairing in Few-Fermion Systems with Attractive Interactions.} \href{https://doi.org/10.1103/PhysRevLett.111.175302}{Phys. Rev. Lett. {\bf 111}, 175302 (2013).}



\bibitem{Gharashi_2015} S. E. Gharashi and D. Blume, {\it Tunneling dynamics of two interacting one-dimensional particles.} \href{https://doi.org/10.1103/PhysRevA.92.033629}{Phys. Rev. A {\bf 92},  033629 (2015).}

\bibitem{Lode_2012} A. U. J. Lode, A. I. Streltsov, K. Sakmann, O. E. Alon, and L. S. Cederbaum, {\it How an interacting many-body system tunnels through a potential barrier to open space.} \href{https://doi.org/10.1103/PhysRevX.8.031042}{Proc. Natl. Acad. Sci. {\bf 109}, 13521 (2012).}

\bibitem{Lundmark_2015} R. Lundmark, C. Forss\'en, and J. Rotureau, {\it Tunneling theory for tunable open quantum systems of ultracold atoms in one-dimensional traps.} \href{https://doi.org/10.1103/PhysRevA.91.041601}{Phys. Rev. A {\bf 91},  041601 (2015).}

\bibitem{Rontani_2012} M. Rontani, {\it Tunneling Theory of Two Interacting Atoms in a Trap.} \href{https://doi.org/10.1103/PhysRevLett.108.115302}{Phys. Rev. Lett. {\bf 108}, 115302 (2012).}

\bibitem{Volosniev_2014} A. G. Volosniev, D. V. Fedorov, A. S. Jensen, M. Valiente, N. T. Zinner, {\it Strongly interacting confined quantum systems in one dimension.} \href{https://doi.org/10.1038/ncomms6300}{Nat. Commun. {\bf 5}, 5300 (2014).}

\bibitem{Ishmukhamedov_2017} I.S. Ishmukhamedov and V.S. Melezhik, {\it Tunneling of two bosonic atoms from a one-dimensional anharmonic trap.} \href{https://doi.org/10.1103/PhysRevA.95.062701}
{Phys. Rev. A {\bf 95}, 062701 (2017).}

\bibitem{Ishmukhamedov_2019} I. Ishmukhamedov and A. Ishmukhamedov, {\it Tunneling of two interacting atoms from excited states.} \href{https://doi.org/10.1016/j.physe.2018.12.026}{Phys. E: Low-dimensional Syst. Nanostruct. {\bf 109}, 24 (2019).}

\bibitem{Zhao_2017} X. Zhao, D. A. Alcala, M. A. McLain, K. Maeda, S. Potnis, R. Ramos, A. M. Steinberg, and L. D. Carr, {\it Macroscopic quantum tunneling escape of Bose-Einstein condensates.} \href{https://doi.org/10.1103/PhysRevA.96.063601}{Phys. Rev. A {\bf 96},  063601 (2017).}

\bibitem{Dobrzyniecki_2018} J. Dobrzyniecki and T. Sowi\ifmmode \acute{n}\else \'{n}\fi{}ski, {\it Dynamics of a few interacting bosons escaping from an open well.} \href{https://doi.org/10.1103/PhysRevA.98.013634}{Phys. Rev. A {\bf 98}, 013634 (2018).}

\bibitem{Brugger_2025} J. Brugger, Ch. Dittel, A. Buchleitner, {\it Two-particle tunneling and the impact of interaction}, \href{https://doi.org/10.1103/PhysRevA.111.033308}{Phys. Rev. A {\bf 111}, 033308 (2025).}


\bibitem{Amirkhanov_1966} I. Amirkhanov and B. N. Zakhariev, {\it Violation of barrier penetration symmetry for composite particles.} \href{http://jetp.ras.ru/cgi-bin/e/index/e/22/4/p764?a=list}{Sov. Phys. JETP {\bf 22}, 764 (1966).}

\bibitem{Bondar_2010} D. I. Bondar, W.-K. Liu, and M. Yu. Ivanov, {\it Enhancement and suppression of tunneling by controlling symmetries of a potential barrier.} \href{https://doi.org/10.1103/PhysRevA.82.052112}{Phys. Rev. A {\bf 82}, 052112 (2010).}

\bibitem{Bilokon_2025} E. Bilokon, V. Bilokon,  D. R. Lindberg, et al. {\it Few-fermion resonant tunneling and underbarrier trapping in asymmetric potentials.} \href{https://doi.org/10.1038/s42005-025-02189-9}{Commun. Phys. {\bf 8}, 259 (2025).}

\bibitem{Gomez_2019} J. P. Gomez, A. Minguzzi, and M. Olshanii, {\it Traces of integrability in scattering of one-dimensional dimers on a barrier.} \href{https://doi.org/10.3390/atoms11050077}{New J. Phys. {\bf 21}, 023008 (2019).}
\bibitem{Olshanii_2025} M. Olshanii, M. Albert, G. Aupetit-Diallo, P. Vignolo, S. G. Jackson,  {\it Exact solution for two $\delta$-interacting bosons on a ring in the presence of a $\delta$-barrier: Asymmetric Bethe Ansatz for spatially odd states.} \href{https://doi.org/10.21468/SciPostPhysCore.8.4.083}{SciPost Phys. Core {\bf 8}, 083 (2025).}


\bibitem{LeBlanc_2007} L. J. LeBlanc and J. H. Thywissen, {\it Species-specific optical lattices.} \href{https://doi.org/10.1103/PhysRevA.75.053612}{Phys. Rev. A {\bf 75}, 053612 (2007).}

\bibitem{Shotter_2011} A. C. Shotter and M. D. Shotter, {\it Quantum mechanical tunneling of composite particle systems: Linkage to sub-barrier nuclear reactions.} \href{https://doi.org/10.1103/PhysRevC.83.054621}{Phys. Rev. C {\bf 83}, 054621 (2011).}

\bibitem{Barker_2020} A. J. Barker, S. Sunami, D. Garrick, A. Beregi, K. Luksch, E. Bentine and C. J. Foot, {\it Realising a species-selective double well with multiple-radiofrequency-dressed potentials.} \href{https://doi.org/10.1088/1361-6455/ab9152}{J. Phys. B: At. Mol. Opt. Phys. {\bf 53}, 155001 (2020).}


\bibitem{Abramowitz} M. Abramowitz and I. Stegun, {\it Handbook of Mathematical Functions with Formulas, Graphs, and Mathematical Tables} (United States Department of Commerce, National Bureau of Standards 1964).

\bibitem{Panochko_2021} G. Panochko and V. Pastukhov, {\it Two- and three-body effective potentials between impurities in ideal BEC.} \href{https://doi.org/10.1088/1751-8121/abdbc5}{J. Phys. A: Math. Theor. {\bf 54}, 085001 (2021).}

\bibitem{Hryhorchak_2023_2} O. Hryhorchak and V. Pastukhov, {\it Trapped Ideal Bose Gas with a Few Heavy Impurities.} \href{https://doi.org/10.3390/atoms11050077}{Atoms {\bf 11}, 77 (2023).}



		
\end{thebibliography}
\end{document}